\documentclass{aastex63}

\usepackage[utf8]{inputenc}
\usepackage[titletoc, title]{appendix}
\usepackage{CJKutf8}
\usepackage{float}
\usepackage{blindtext}
\usepackage{gensymb}
\usepackage{amsmath}
\usepackage{mathtools}

\def \MIDAS {Michigan Institute for Data Science, University of Michigan, Ann Arbor, MI 48109, USA}
\def \Physics {Department of Physics, University of Michigan, Ann Arbor, MI 48109, USA}
\def \CfA {Center for Astrophysics | Harvard \& Smithsonian, 60 Garden Street, Cambridge, MA 02138, USA}

\begin{document}
\shorttitle{A Novel Orbit Parameterization in Spherical Coordinates}
\shortauthors{Napier and Holman}

\submitjournal{PSJ}

\title{A Novel Orbit Parameterization in Spherical Coordinates}

\correspondingauthor{Kevin J. Napier}
\email{kjnapier@umich.edu}
\author[0000-0003-4827-5049]{Kevin~J.~Napier}
\affiliation{\MIDAS}
\affiliation{\Physics}
\affiliation{\CfA}

\author[0000-0002-1139-4880]{Matthew~J.~Holman}
\affiliation{\CfA}

\begin{abstract}
We present a novel orbit parameterization in spherical coordinates. This parameterization enables the mixing of varying and invariant orbital parameters, and clarifies the physics of the orbit. It also simplifies the process of placing synthetic populations at exactly specified locations on the sky, which is particularly useful for survey design and simulation studies.
\end{abstract}

 
\section{Introduction}
\label{sec:intro}




For centuries scientists have used Keplerian elements to describe orbital motion \citep{Kepler1609}. The elements are convenient because they are mostly invariant---a property that was especially important before computers were as powerful as they are today. There are several useful variations of Keplerian elements, including the Delaunay coordinates and the canonical Poincare coordinates~\citep{Murray1999}. While the Keplerian elements are incredibly useful, they tend to obfuscate a lot of information about an orbit. For example, it is usually not easy to determine the barycentric distance or position on the sphere by inspecting the elements. Cartesian position and velocity vectors avoid this problem, but they have the opposite drawback of obscuring the physics of the orbit. In many cases it would be useful to work with a combination of orbital parameters that combine the advantages of both approaches.

In this paper we introduce a new orbit parameterization in spherical coordinates that makes it simple to represent orbits with a variety of invariant and varying parameters. We define the coordinate system in Section \ref{sec:coordinates} and highlight some of its properties in Section \ref{sec:properties}. We show example use-cases in Sections \ref{sec:encounters} and \ref{sec:orbit-divergence}, and we conclude with a brief discussion in Section \ref{sec:discussion}.

\section{Coordinate System Definition}
\label{sec:coordinates}

For this parameterization we work in heliocentric (or barycentric) spherical coordinates with an arbitrary reference plane. First we define a unit vector that specifies the position of a body on the heliocentric unit sphere as
\begin{equation}
    \mathbf{\hat{r}} = \langle \cos{\theta}\cos{\varphi}, \cos{\theta}\sin{\varphi}, \sin{\theta} \rangle,
    \label{eq:pointing}
\end{equation}
where $\varphi$ is the body's longitude and $\theta$ is its latitude as measured from the equator. We then define two more unit vectors that are mutually orthogonal to $\mathbf{\hat{r}}$,
\begin{align}
    \mathbf{\hat{A}} &= \langle -\sin{\varphi}, \cos{\varphi}, 0 \rangle
    \\
    \mathbf{\hat{D}} &= \langle -\sin{\theta}\cos{\varphi}, -\sin{\theta}\sin{\varphi}, \cos{\theta} \rangle,
    \label{eq:AD}
\end{align}
thus specifying a complete right-handed orthogonal basis~\citep{danby}. 

\begin{figure}[H]
    \centering
    \includegraphics[width=0.5\textwidth]{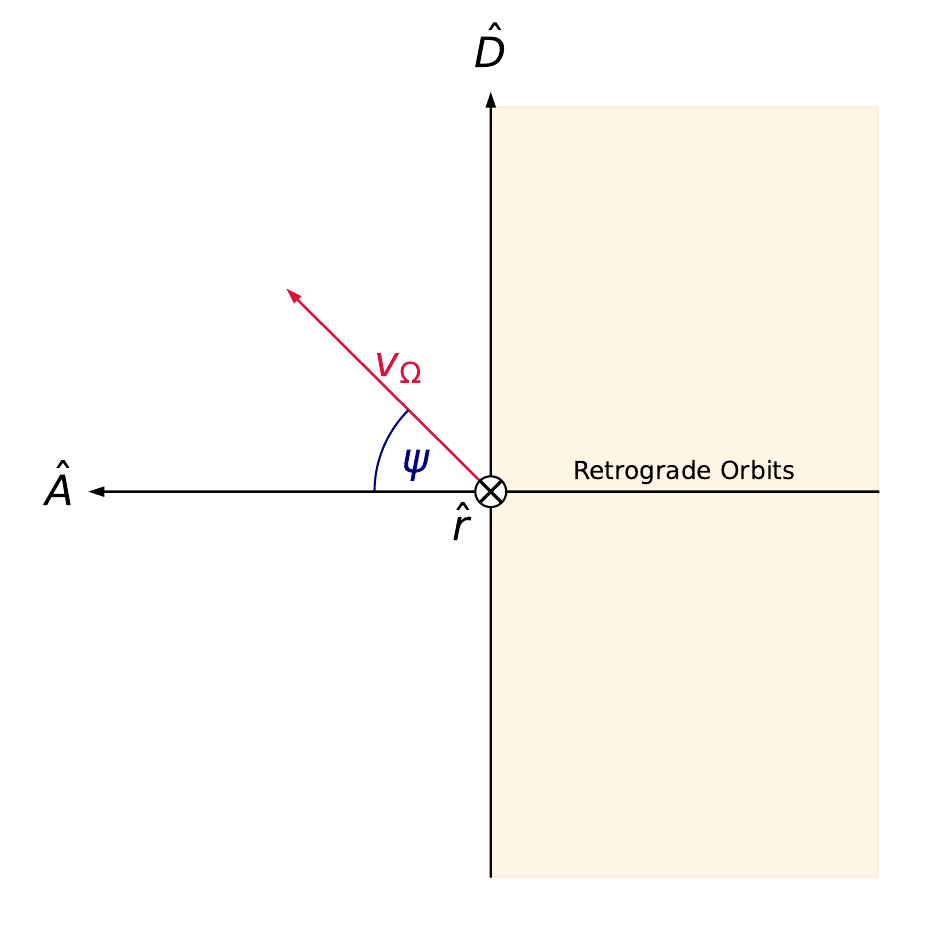}
    \caption{Schematic of the coordinate system described in this paper. The tangential coordinate plane is defined by the $\mathbf{\hat{A}}$ and $\mathbf{\hat{D}}$ vectors. The $\mathbf{\hat{r}}$ vector is going into the page at the coordinate origin. The angle $\psi$ is measured with respect to the $\mathbf{\hat{A}}$ vector, and $v_\Omega$ gives the body's speed in the plane. Note that for values of $\psi \in (\pi/2, 3\pi/2)$, all orbits are retrograde with respect to the sphere's equator.}
    \label{fig:coordinates}
\end{figure}

We define the body's distance and its radial velocity as $r$ and $v_r$, respectively. Finally, we define the body's tangential speed as $v_{\Omega}$ and its direction as the angle $\psi$, which is measured with respect to $\mathbf{\hat{A}}$ (see Figure \ref{fig:coordinates}). The state vector of the orbit can then be obtained as follows.
\begin{align}
    \mathbf{r} &= r \mathbf{\hat{r}}
    \label{eq:position} \\
    \mathbf{v} &= v_r \mathbf{\hat{r}} + v_{\Omega} \left( \cos{\psi} \mathbf{\hat{A}} + \sin{\psi} \mathbf{\hat{D}}\right)
    \label{eq:velocity}
\end{align}
Thus we have a complete specification of an orbit using the parameters $\langle \varphi, \theta, r, v_r, v_\Omega, \psi \rangle$.

\section{Properties}
\label{sec:properties}

\subsection{Conserved Quantities}
\label{subsec:conserved-quantities}

The coordinate system introduced in Section \ref{sec:coordinates} has the nice property that all of the conserved orbital quantities can be expressed as terse combinations of its parameters. An orbit's specific energy is given by
\begin{equation}
    \epsilon = \frac{v_r^2 + v_\Omega^2}{2} - \frac{\mu}{r},
    \label{eq:epsilon}
\end{equation}
and its specific angular momentum is given by
\begin{equation}
    h = r v_\Omega.
    \label{eq:h}
\end{equation}
It is also simple to show that the $z$ component of its angular momentum is given by
\begin{equation}
    h_z = r v_\Omega \cos{\theta} \cos{\psi}.
    \label{eq:hz}
\end{equation}

\subsection{Hybrid Representations}
\label{subsec:hybrid-representations}

This basis also enables mixed representations by Keplerian and spherical elements, which can be useful for solving certain classes of problems. There is a bijection between $\langle r, v_r, v_\Omega \rangle$ and $\langle \epsilon, h, r \rangle$, making it is easy to express some of the Keplerian elements using the spherical basis parameters. For example, the semi-major axis $a$, eccentricity, $e$, inclination, $i$, and true anomaly, $f$, can be expressed as follows.\footnote{It is also simple to express the true anomaly's rate of change as $\dot{f} = v_\Omega / r$.}
\begin{align}
    a &= -\frac{\mu}{2 \epsilon} \\[1em]
    e &= \sqrt{1 + \frac{2 \epsilon h^2}{\mu^2}} \\[1em]
    \cos{i} &= \frac{h_z}{h} = \cos{\theta} \cos{\psi} \\[1em]
    \cos{f} &= \frac{1}{e} \left( \frac{h^2}{\mu r} - 1 \right)
\end{align}
From the above equations we can see that $\langle \varphi, \theta, a, e, f, \psi \rangle$ is a valid orbit representation with three varying spherical parameters, two invariant Keplerian elements, and one always-known Keplerian element. This representation is particularly useful for orbit linking and image stacking, which we will explore in depth in future work.

It is also possible to use the orbital inclination $i$ as an orbital parameter, rather than $\psi$. However, since $\cos{i} = \cos{\theta} \cos{\psi}$, there are two unique values of $\psi$ that correspond to the given inclination, both with the same magnitude, but one positive and one negative. If $\psi$ is positive, the orbit is ascending, and \textit{vice versa}. This limitation means that the inclination on its own is not a suitable substitute for $\psi$. However, we can alleviate this issue by introducing a new parameter, $\kappa$, which can take on discrete values of $\pm 1$, informing the sign of $\psi$. Then we can parameterize an orbit using the set of parameters $\langle \varphi, \theta, a, e, f, i, \kappa \rangle$, as long as we obey the constraint $\theta < i$ (because a body's latitude can never exceed its inclination). This parameterization has effectively separated the Keplerian invariants from the orbit's position on the sphere.

Finally, we note that it is possible to parameterize the orbit using $\langle a, e, r \rangle$ rather than $\langle a, e, f \rangle$. However, in doing so we lose information about whether the body is moving toward or away from the coordinate origin. To alleviate this issue, we introduce the parameter $\iota$, which can take on discrete values of $\pm 1$, informing the sign of $v_r$. Thus, it is possible to specify an orbit using the parameters $\langle \varphi, \theta, r, a, e, i, \kappa, \iota \rangle$. This parameterization allows one to place a body at an exact location in space, with any desired values of $a$, $e$, and $i$ (as long as the elements are commensurate with $r$ and $\theta$). The four-fold multiplicity introduced by $\kappa$ and $\iota$ is a small price to pay for the leverage gained by selecting a position in space, as well as three invariants.

\section{Example: Lucy Encounters}
\label{sec:encounters}
Consider that we have a spacecraft on a pre-determined trajectory, and we want to discover a body for it to observe at close range. Where and when should we search for the target? One way to approach the problem is by generating a synthetic population of encounterable bodies and calculating their sky positions from some location on Earth at various epochs to see if the objects cluster in some location on the sky. We can ensure that the synthetic bodies are encounterable by assigning them values of $\langle \varphi, \theta, r \rangle$ and epochs identical to those of the spacecraft's nominal trajectory. Then we can choose the objects' $\langle a, e, i, \kappa, \iota \rangle$ values consistent with some realistic orbital distribution. 

\begin{figure}[H]
    \centering
    \includegraphics[width=0.9\textwidth]{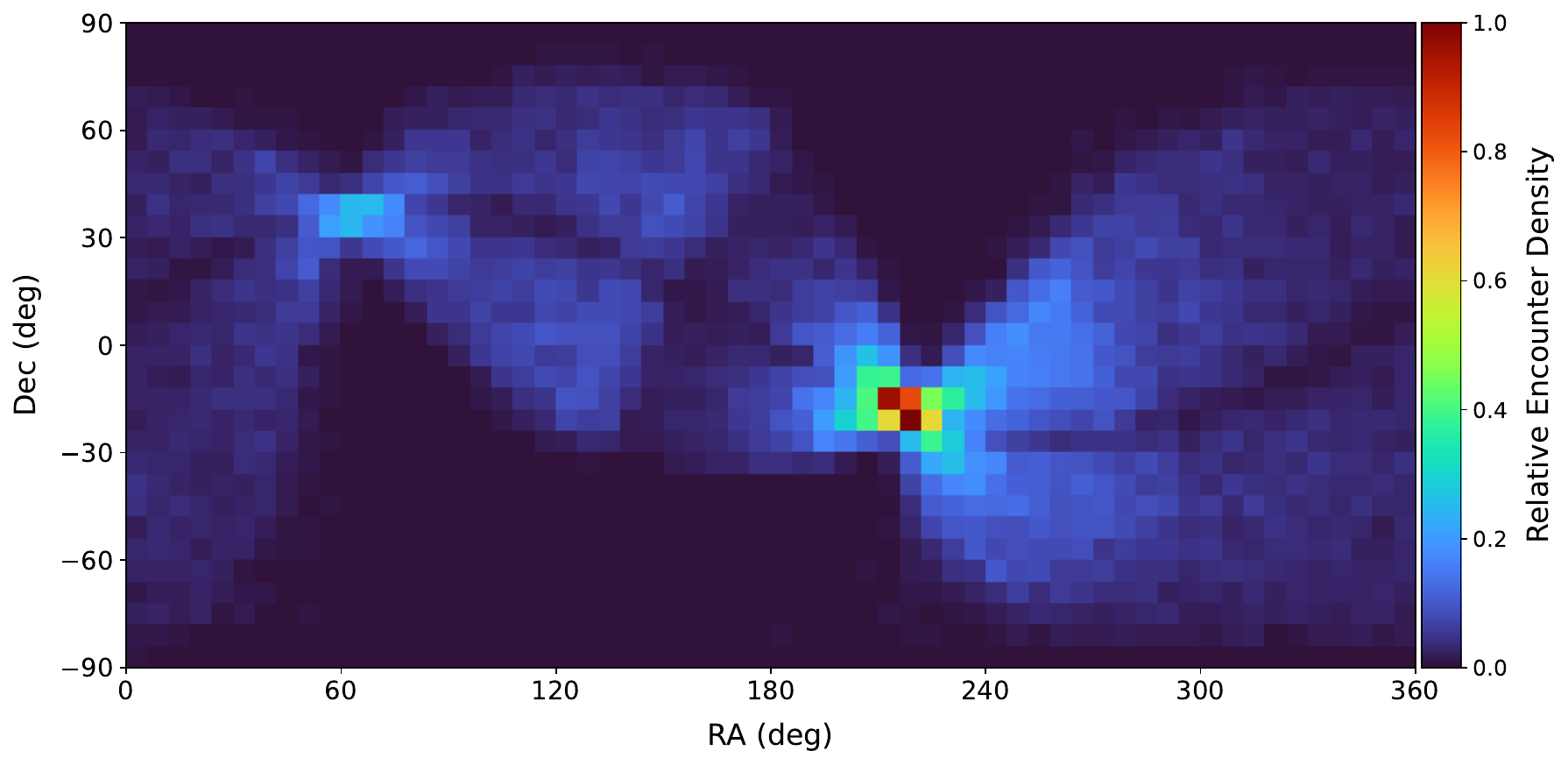}
    \caption{Relative sky density of synthetic objects encounterable by Lucy between 16 July 2032 and 15 January 2033, as viewed from Cerro Tololo on July 16 2024.}
    \label{fig:lucy-encounter}
\end{figure}

For this demonstration, we use the following procedure to generate encounterable objects for Lucy. We note that our chosen distributions of the Keplerian elements are not realistic, but they are fine for the present exercise. 
\begin{enumerate}
    \item Query Lucy's $\langle r, \theta, \varphi \rangle$ at a random epoch $t$ between 16 July 2032 and 15 January 2033.
    \item Generate a random semi-major axis value, $a \in [2.4, 5.4]$ au.
    \item Generate a random pericenter distance, $q \in [1.4, a]$ au.
    \item Generate a random inclination, $i \in [|\theta|, 60]$ deg. 
    \item Randomly choose $\kappa \in \{-1, 1\}$.
    \item Randomly choose $\iota \in \{-1, 1\}$.
    \item Calculate $\langle v_r, v_\Omega, \psi \rangle$ from $\langle \theta, a, e, i, \kappa, \iota \rangle$.
    \item Record $\langle r, \theta, \varphi, v_r, v_\Omega, \psi, t \rangle$ in a list of encounterable objects.
\end{enumerate}
We repeat the above procedure until we have accumulated 50000 encounterable objects. Finally, we calculate the objects' apparent sky positions as viewed from Cerro Tololo on July 16 2024. It is clear from Figure \ref{fig:lucy-encounter} that there is a preferred location on the sky to maximize the likelihood of finding an encounterable object. Of course there are many ways that this calculation can be improved, perhaps by choosing more realistic orbital distributions, or a variety of different encounter or observation dates. Nevertheless, this procedure is an efficient way of carrying out the experiment.

\section{Example: Orbit Divergence}
\label{sec:orbit-divergence}

In general we can calculate the heliocentric angular separation $\xi(t)$ between two orbits as
\begin{equation}
    \xi(t) = \arccos{\left(\frac{\mathbf{r}_1(t) \cdot \mathbf{r}_2(t)}{r_1(t) r_2(t)}\right)}.
    \label{eq:separation}
\end{equation}
If we assume two-body motion we can express a body's position as a function of time as
\begin{equation}
    \mathbf{r}(t) = \mathbf{r}(t_0) f(t, t_0) + \mathbf{v}(t_0) g(t, t_0).
\end{equation}
Here the so-called Gauss $f$ and $g$ functions are expressed for bound orbits as
\begin{align}
    f(t, t_0) &= 1 - \frac{a}{r(t_0)} \left[ 1 - \cos(\Delta E) \right] \\
    g(t, t_0) &= (t - t_0) + n^{-1} \left[ \sin(\Delta E) - \Delta E \right]
\end{align}
where $a$ is the semi-major axis, $n$ is the mean motion, $E$ is the eccentric anomaly, and $\Delta E$ is the change in the eccentric anomaly~\citep{danby}. We define the state vectors for two orbits as $\langle \mathbf{r}_1, \mathbf{v}_1 \rangle$ and $\langle \mathbf{r}_2, \mathbf{v}_2 \rangle$, so that we can write 
\begin{align*}
    \mathbf{r}_1(t) &= \mathbf{r}_1(t_0) f_1(t, t_0) + \mathbf{v}_1(t_0) g_1(t, t_0) \\
    \mathbf{r}_2(t) &= \mathbf{r}_2(t_0) f_2(t, t_0) + \mathbf{v}_2(t_0) g_2(t, t_0).
\end{align*}

Now consider the special case where we have a well-constrained value of an orbit's $\langle \varphi, \theta \rangle$, and we want to explore how perturbations to the other orbital parameters affect the divergence of the orbit's trajectory in $\langle \varphi, \theta \rangle$ over time. It is convenient in this case to use the pure form of the spherical orbit parameterization, $\langle \varphi, \theta, r, v_r, v_\Omega, \psi \rangle$. We consider two orbits with identical values of $\langle \varphi, \theta \rangle$, but arbitrary values of $\langle r, v_r, v_\Omega, \psi \rangle$. Since both orbits have the same $\langle \varphi, \theta \rangle$ coordinates, they have the same $\mathbf{\hat{r}}$, $\mathbf{\hat{A}}$, and $\mathbf{\hat{D}}$. Then Equation (\ref{eq:separation}) can be expressed as
\begin{equation}
    \begin{split}
    r_1(t) r_2(t) \cos \xi(t) &= r_1(t_0) r_{2}(t_0) f_1(t, t_0) f_2(t, t_0) \\
    &+ r_{1}(t_0) v_{r 2}(t_0) f_1(t, t_0) g_2(t, t_0) \\
    &+ r_{2}(t_0) v_{r 1}(t_0) f_2(t, t_0) g_1(t, t_0) \\
    &+ g_1(t, t_0) g_2(t, t_0) \left\{ v_{r 1}(t_0) v_{r 2}(t_0) + v_{\Omega 1}(t_0) v_{\Omega 2}(t_0) \cos{\left[\psi_{1}(t_0) - \psi_{2}(t_0)\right]} \right\}.
    \end{split}
    \label{eq:expanded-separation}
\end{equation}
All of the values on the right-hand side of Equation (\ref{eq:expanded-separation}) are either known or easily computed, and $r_1(t)$ and $r_2(t)$ can be calculated using Kepler's Equation. Thus the equation represents an analytic solution of the angular separation between the two bodies as viewed from the barycenter at all times. In many cases this expression serves as a good approximation for the angular separation between the bodies as viewed from Earth; the fidelity of the approximation improves with increasing distance. This result is particularly useful for orbit linking applications such as \texttt{HelioLinC} \citep{holman2018}, which we will further explore in future work.

\section{Discussion}
\label{sec:discussion}

In this paper we have introduced a novel orbit parameterization using spherical coordinates. The parameterization enables the mixing of varying and invariant orbital parameters.  We have traded the invariant elements longitude of pericenter ($\varpi$) and longitude of ascending node ($\Omega$) for a latitude and a longitude. Additionally, the this basis has the nice property that it allows physical quantities to be expressed as terse arithmetic combinations of scalars, clarifying the physics of the orbit.

We can already identify several use cases for this parameterization. For example, this formulation makes it trivial to place synthetic populations at exactly specified locations in space, with any chosen distributions in some of the canonical Keplerian elements. This feature is particularly desirable for survey design and simulation studies.

In some ways, this parameterization is related to that introduced by \citet{Bernstein2000}. It has the advantage, however, that it applies generally to the whole sphere, while the parameterization by \citet{Bernstein2000} makes use of a tangent plane. This property makes our new parameterization particularly useful for minor planet linking and image stacking, which we explore in forthcoming work.

\acknowledgments
This project is supported by Schmidt Sciences, LLC.  M.J.H. gratefully acknowledges support from the NSF (grant No. AST-2206194), the NASA YORPD Program (grant No. 80NSSC22K0239), and the Smithsonian Scholarly Studies Program (2022, 2023).

\bibliographystyle{aasjournal}
\bibliography{references}

\end{document}